# Optimization and AMS Modeling for Design of an Electrostatic Vibration Energy Harvester's Conditioning Circuit with an Auto-Adaptive Process to the External Vibration Changes

[1]Dimitri Galayko, [2]Philippe Basset, [2]Ayyaz Mahmood Paracha
[1]LIP6, Paris, France, ESYCOM, [2]Noisy-le-Grand, France
Contact : dimitri.galayko@lip6.fr

*Abstract*-This paper presents an analysis and system-level design of a capacitive harvester of vibration energy composed from a mechanical resonator, capacitive transducer and a conditioning circuit based on the BUCK DC-DC converter architecture. The goal of the study is to identify optimal power performance of the system, to understand the electromechanical coupling phenomena and to propose the optimal timing of switching between charge pump and flyback circuits. To achieve the study we provided a VHDL-AMS/ELDO mixed model based on physical equations describing the resonator and transducer operation. To test different algorithms of the switching, we developed a behavioral functional model of the switch commuting between the two operation phases. We demonstrated that to guarantee an optimal power generation, switching should be driven by the internal state of the circuit. This paper provides the keys of the underlying analysis and provides a basic algorithm of the switch "intelligent" command.

## I. Introduction

With evolution of microelectronic technologies, the miniaturization and integration of electronic systems will be pursued. The microelectronic industry is moving toward very compact, autonomous and pervasive systems which use multi-physics signal processing, e.g, embedded sensors and sensor networks. However, the energy supply autonomy of such systems remains the key problem, and on of the solutions can be provided by harvesting energy from the environment. Recently a keen interest has been demonstrated toward harvesting of energy of environment vibrations, being motivated by applications of sensor network in moving objects (transportation…).

A mechanical energy harvester is composed from a mechanical resonator, an electromechanical transducer and a conditioning circuit achieving the energy transfer from the transducer toward the electrical load. In this paper, we deal with a capacitive (electrostatic) transducer. Such systems are attractive for two reasons : firstly, they can be fabricated in silicon micromachined process compatible with CMOS clean rooms, secondly, in difference with piezoelectric-based systems, they are more likely to provide a large frequency passband.

However, study and design of a capacitive harvester are challenging tasks, since they require a design of a complex conditioning circuit managing the electrical charge flow on the transducer capacitor, and an analysis of coupled electromechanical behavior. For this reason a reliable model of the system is needed for the design.

The object of our study it the circuit presented in fig. 1, firstly introduced in [1]. This circuit is composed from a charge pump and a flyback circuit. The idea of energy harvesting is resumed in three steps : putting an electrical charge on a variable capacitor when the capacitance is high, reducing the capacitance thanks to the mechanical vibrations, discharging the capacitor. Following the formula $W=Q^2/(2C)$, the discharge energy is higher than the energy spent to charge the capacitor.

However, a practical implementation requires a more complex harvesting cycle, related to the need of automatic charging and discharging of the variable capacitor which is achieved by a charge pump.

## II. Charge Pump Operation

The charge pump is composed of three capacitors : a large-capacitance $C_{res}$ (~1 µF), a smaller $C_{store}$ (~1 nF) and $C_{var}$ which is the variable transducer capacitance and which varies typically between several tens and several hundreds of picofarads. $C_{res}$ capacitor is supposed to be big, so to keep a constant voltage during all phases of the system operation.

The role of the charge pump is to transfer the charges from $C_{res}$ toward $C_{store}$. Since $C_{store} < C_{res}$, this transfer requires an external energy: the latter is provided from the mechanical domain through the variations of the capacitance $C_{var}$ (details can be found in [1]). The energy harvested during each pump is accumulated in the difference between the voltages of $C_{res}$ and $C_{store}$ capacitors, or in other words, in the energy of an equivalent capacitor composed from series connection of $C_{res}$ and $C_{store}$ and charged to a voltage equal to the difference between $C_{res}$ and $C_{store}$ voltage :

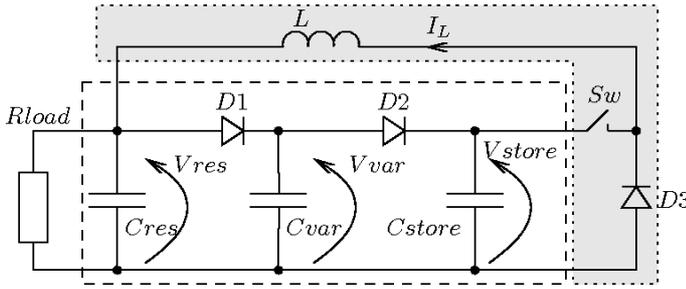

Figure 1. Circuit of the energy harvester. In white : charge pump, in gray : flyback circuit

$$\Delta W_{1 \to n} = \frac{C_{res} C_{store}}{2(C_{res} + C_{store})}(Vstore_n - Vres)^2 \approx \frac{C_{store}}{2}(Vstore_n - Vres)^2 \quad (1)$$

where $\Delta W_{1 \to n}$ is the energy accumulated by the pump charge during n pump cycles, $Vstore_n$ is Cstore voltage at the end of the $n^{th}$ cycle.

Observing the QV diagram of the charge pump operation (fig. 2), one can see that the curve describing the cycle is changing each cycle. Indeed, if at the beginning (point 1) Vres=Vvar=Vstore=$V_0$, and Cvar=Cmax, the first cycle will be triangular (black). However, the second cycle will contain an addidional point (1bis), thus will be larger and less high that the first one, and will be a trapeze (red). It is possible to see that at cycle n the point 1bis is situated at V=Vstore_1, and that at each cycle Vstore increases, thus, the cycle becomes larger. At the same time, the minimal charge of the variable capacitor increases, so the cycle becomes thinner (the line 1→2 is always situated at fixed Q-coordinate equal to CmaxV_0). From the left and from the right sides, the QV diagram is bounded by the lines corresponding to Cmin and Cmax values of the variable capacitance : Q=CminV and Q=CmaxV. Hence the lines Q=CminV and Q= CmaxV_0 (line 1→1bis ) fix the maximal value of Vstore, which is equal to V_0Cmax/Cmin and called "saturation voltage". When Vstore is close to Vstore_{sat}, the cycle is very thin, and at n→∞ it degenerates to a line.

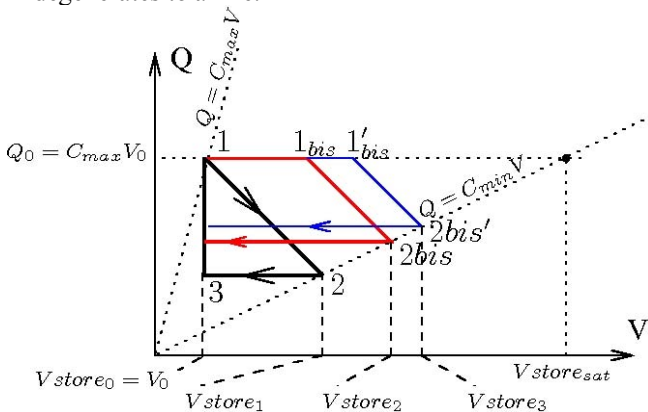

Figure 2. QV diagram of the charge pump operation

It is known that the area of the a clockwise QV cycle is numerically equal to the harvested energy. Visually, it is possible to see that the first cycle (black) has a smaller area than the second (red), but starting from some n the cycle areas are tending to zero. This tendency is also demonstrated in the plots on the fig. 3. Hence, there is a cycle number at which the pump harvests the maximal amount of energy. Supposing that the vibrations are periodical, and each pump cycle takes exactly the same time, one can see that there is an operating mode at which the pump produce a maximum power. This is a theoretical physical limit of the charge pump performance, for the given capacitance values, ideal diodes and given $V_0$ voltage, and it is very important to know it.

### III   ESTIMATION OF THE MAXIMAL POWER.

Here we will give the guidelines of the estimation of the theoretical maximal power.

The problem is resumed in the following way. It is required to find the maximum of the function :

$$\Delta W(n) = \Delta W_{1 \to n} - \Delta W_{1 \to n-1}, \quad (2)$$

Where $\Delta W_{1 \to n}$ is determined by the formula (1).

The formula for the value of the voltage Vstoren was derived in [1] :

$$Vstore_n = V_0 \left[ \left(1 - \frac{C\max}{C\min}\right) \left(\frac{C\max}{Cstore + C\max}\right)^n + \frac{C\max}{C\min} \right] \quad (3)$$

Looking for the maximum of (2) by n (using (1) and (3)), it is possible to find the number n of the most "efficient" cycle. It is impossible to find n in closed form, so, a numeric analysis is necessary.

However, it is possible to have an intuitive idea from the following considerations.

If Cmin << Cmax, for low n, the evolution of Vstore is linear :

$$Vstore_n \approx V_0 \left[ 1 + \frac{C\max}{Cstore} \cdot n \right] \quad (4)$$

(this formula is obtained by limited development in power series).

Using (4), it is possible to express $\Delta W(n)$ :

$$\Delta W(n) = \frac{C\max V_0^2}{2} \frac{C\max}{Cstore}(2n - 1) \quad (5)$$

This function has no maximum under the hypothesis which have been made. However, in practice (when Cmin is non-zero), there will be a saturation of Vstore, and from some n and the corresponding Vstore $\Delta W(n)$ will start to decrease and will become zero when n is infinite (fig. 3).

This reasoning proves that there is a one and only one cycle of charge pump for which the latter generate the maximum power. The very important conclusion is that the index of this cycle, and more generally, the optimal operation zone of the pump charge, can be defined using only information about Vstore.

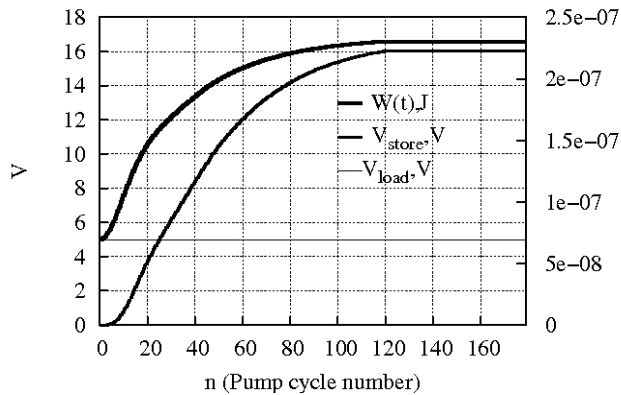

Figure 3. Harvested energy of the pump charge and the voltages on Cstore and Cres capacitors versus the number of the pump cycle.

### IV ENSURING AN OPTIMAL CHARGE PUMP OPERATION

As showed the preceding demonstration, the charge pump operation is not symmetric in time, and only during some limited time interval the pump generates a maximal power. To guarantee a continuous power generation, a flyback circuit is needed. Its role is twofold : firstly, to return the charge pump to some earlier state (not obviously initial where Vstore=$V_0$) so to keep its operation in the optimal zone, secondly, to extract the harvested energy from the charge pump and to provide it to the load.

This is achieved with the flyback circuit whose architecture has been inspired from the BUCK DC-DC converter architecture (fig. 1). It composed from the diode D3 and the inductor L. The principle of operating is the following.

When the switch becomes on, the LC circuit CresCstoreL (in which the two capacitors are connected in series) starts to oscillate with the following initial conditions: $I_L=0$, $V_C$=Vstore-Vres. The (composed) capacitor discharges, and the inductor accumulated the energy in its magnetic field. When the difference between Vstore and Vres decreases sufficiently to return the charge pump at the beginning of the optimal operation zone, the switch opens and the inductor discharges on Cres capacitance providing to it electrical charges extracted from the ground. This last phase allows to compensate the loss of the charges of the circuit due to the load and to the parasitic leakage.

From the above analysis of the charge pump, an ideal circuit operation would be such that the charge pump achieves only the cycle corresponding to the optimal n and Vstore, and, at the end of this cycle the flyback circuit return the pump at the state corresponding to the beginning of this cycle.

However, such operation would be ideal only in theory. In practice some losses are always associated with the flyback operation (losses in inductor, switching power etc.). For this reason, the frequency of the flyback circuit activation should be as low as possible. From the other side, enlarging the operating zone of the charge pump to several cycles leads to degradation of the harvested power. These two opposite factors provide and optimal (compromise) definition of the optimal operation zone of the charge pump : the start and end Cres voltages which we call $V_1$ and $V_2$ corresponding to some $n_1$ and $n_2$. Only numeric optimization can provide the exact values given the exponential nature of the corresponding equations. The average harvested power to be optimized is given by :

$$P(n_1, n_2) = \frac{\Delta W_{1 \to n_2} - \Delta W_{1 \to n_1} - Pswitch}{T(n_2 - n_1)}, \quad (6)$$

where Pswitch is the energy loss associated with each switching.

The maximum of this two-variable function is to be found numerically. An example of the corresponding surface is given in fig. 4. The resulting system operation is presented in the diagram of fig. 5.

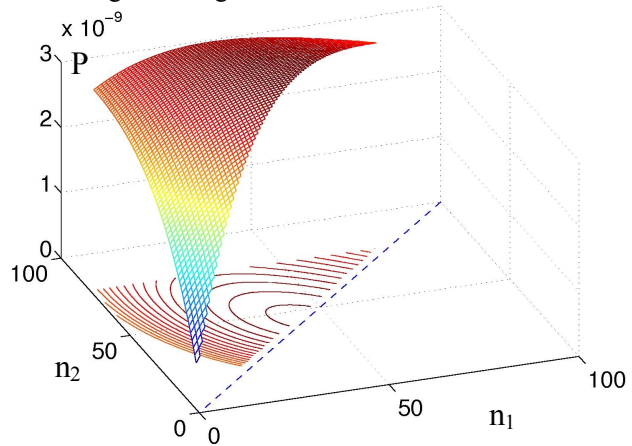

Figure 4 Example of the surface shape P(n1, n2) for the harvester of fig. 1 with component parameters given in below.

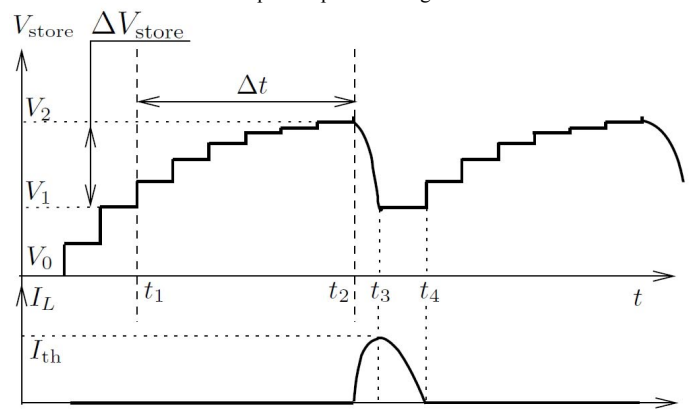

Figure 5 Diagrams demonstrating an optimal circuit operation on a limited zone Vstore(n)

### V OPTIMAL SWITCH OPERATION AND SWITCH FONCTIONAL MODEL

The above considerations suggest the following switching algorithm : the switch should become on (activate the flyback) when Vstore reaches the $V_2$ value corresponding to the right limit of the optimal operation zone, and become off when Vstore decreases to $V_1$ corresponding to the left limit of this zone. In addition, the switch should remember its current state, hence it should be a finite automaton with one-bit memory (fig. 6).

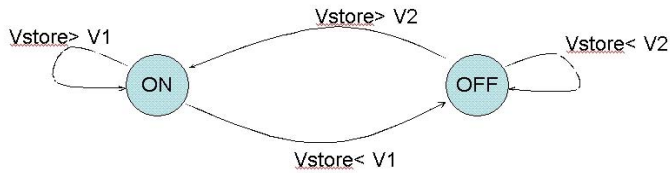

Figure 6 Functional diagram of the switch. $V_1$, $V_2$ are the limits values of the Vstore voltage (fig. 5)

The parameters $V_1$ and $V_2$ are found from the maximization of (6). In this way, switching is entirely driven by the internal state of the circuit. Under the hypothesis of the constant amplitude of the capacitance variation (i.e., fixed Cmax and Cmin), the optimal power generation level is guaranteed with any variation in the external vibration timing.

In this context, the switch is a complex "intelligent" device driven by Vstore voltage and internal algorithm.

### VI    MODEL OF THE HARVESTER

The system of fig. 1 was modeled using a physical VHDL-AMS model for the electromechanical transducer and the resonator, using an ELDO model for the electrical part, and using a functional VHDL-AMS model for he switch [2].

The model of the resonator and the transducer was obtained using the measurement results of a real device we presented in [3] (fig. 7). For this device, fig. 8 presents the variation of the transducer capacitance versus proof mass displacement. This plot provides the information about max-to-min ratio of the transducer capacitance, i.e., about 3.5.

The whole system was modeled in CADENCE environement using AdvanceMS mixed simulatoir. Fig. 9 provides the schematic view of the system.

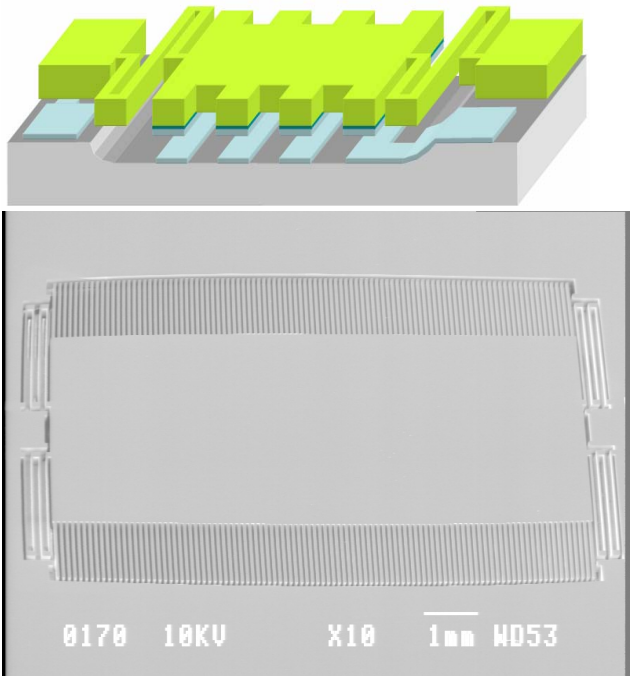

Figure 7. Geometry and the SEM picture of the modeled electromechanical device [3].

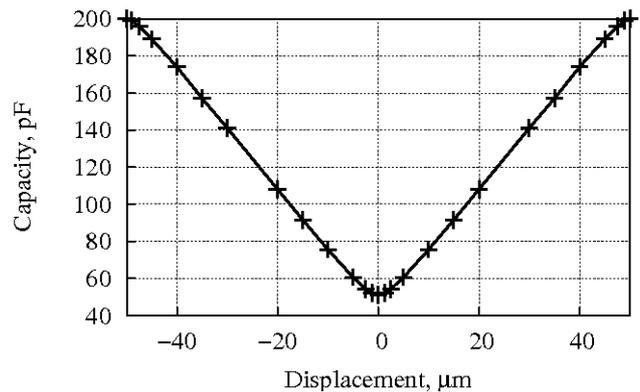

Figure 8. Capacitance versus proof mass displacement for the resonator used in the model.

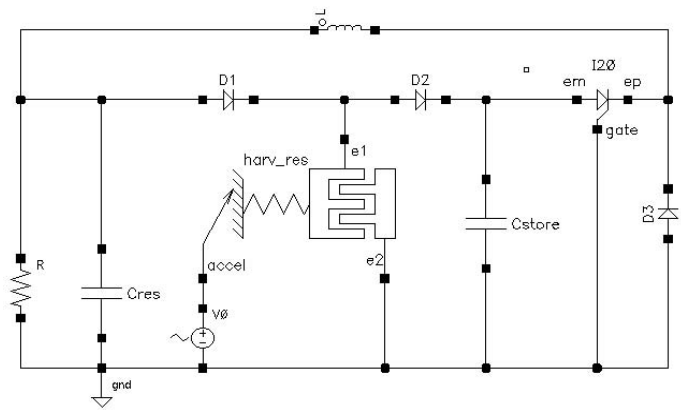

Figure 9. Schematic view of the harvester modeled in CADENCE environment. Cres=1 µF, Cstore=3.3 nF, L=2.5 µH, magnitude of external acceleration ($V_0$ voltage source) 10 ms$^{-2}$, external vibration frequency 298Hz. The mass, elastic and damping constant of the resonator model were respectively of 46.0·10-6 kg, 152.6 Nm$^{-1}$ and 2.19·10-3 Nsm$^{-1}$. The starting voltage of the charge pump (Vres voltage) was set to 5 V.

### VII    SIMULATION RESULTS

The plots of the fig. 10 demonstrates an autonomous operation of the harvester. The voltages $V_1$ and $V_2$ in the switch functional mode were set to 6.5 and 13 V, the starting voltage ($V_0$) being set to 5 V. The simulated plots are in very good adequacy with the theoretical analysis. From these plots a very interesting phenomenon can be observed. One can notice that when the Vstore voltage is at its minimum (just after the flyback phases), the amplitude of vibrations increases slightly. This is a consequence of a complex non-linear coupling between mechanical and electrical domains, and a demonstration of the fundamental law of the conservation of the total energy of the global system.

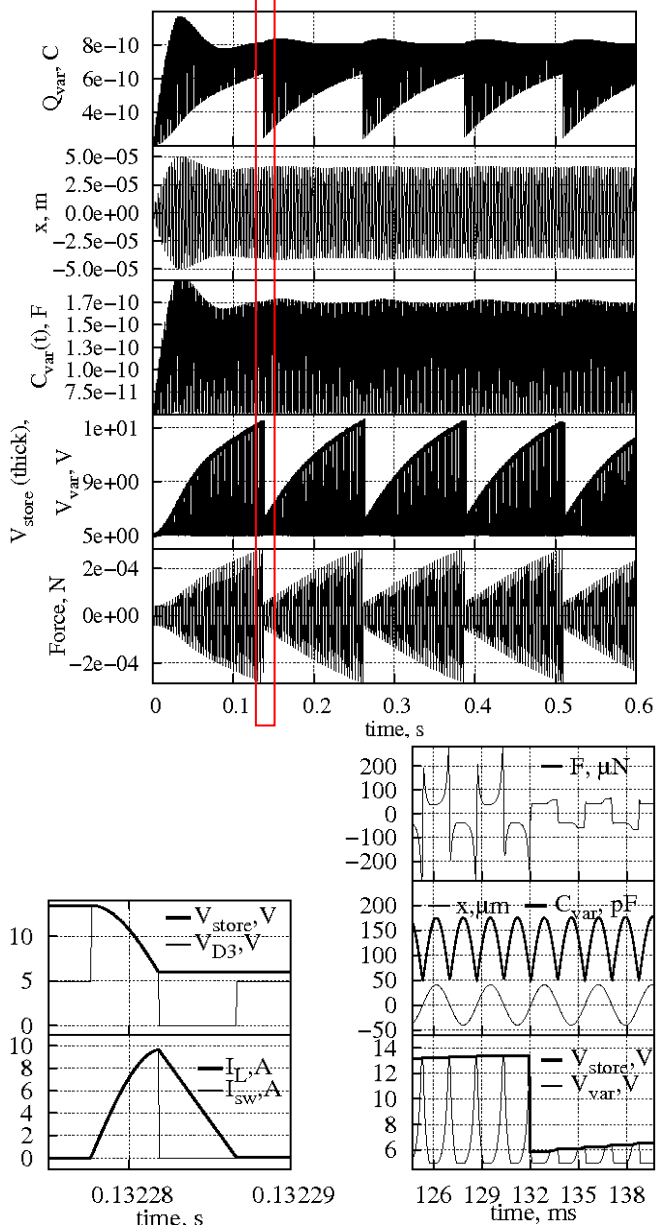

Figure 10. Simulation results. At top: global view of evolution of mechanical and electrical values of the harvester, at bottom left : zoom on the flyback circuit operation, at bottom right : zoom on the charge pump operating before and after a flyback phase. Qvar : charge on the transducer, x : displacement of the proof mass, Force and F : electrical force generated by the transducer, $V_{D3}$: voltage on the diode D3

Indeed, when the voltage of Cstore is high, during the motion from Cmax to Cmin, the transducer must cross the potential barrier from Vres (nearly constant and equal to $V_0$) to Vstore : this appeals a relatively big amount of energy from the mechanical domain. This phenomenon appears as an electrostatic force generated by the transducer which tends to slow down the mass motion. Thus, the higher is Vstore, the lower is the amplitude of motions.

This observation highlights inherent limits of reasoning in purely electrical or mechanical domain. Indeed, nearly all analysis of the pump charge operation are based on the hypothesis of a constant-amplitude transducer capacitance variation. However, this is only valid if the energy stored in the mechanical domain is much higher that the energy stored in electrical domain (which result, actually, in undersized, non-optimized harvesting), and thus, the amplitude of vibrations is not perturbed by the variation of the electrical force. However, when one is interested by harvesting a maximal energy from the mechanical domain, mechanical and electrical energy levels are comparable, and the mentioned coupling phenomenon cannot be neglected : a more complex analysis is needed to estimate the system performance. To demonstrate it in a very simple way, suppose Cmin is zero or very small (which is physically not impossible), thus, from the formula (5) there is no upper limit for the harvested energy ! Obviously, if a realistic mechanical system is present, starting from some Vstore the vibration amplitude will decrease, Cmin-Cmax range will reduce, invalidating the formula.

## VIII CONCLUSIONS AND PERSPECTIVES.

Harvesting of vibration energy using capacitive transducer is a very interesting engineer and scientific problem. In this paper we proposed an approach of synchronous feedback command of the harvesting phases. We proposed a methodology of maximization of the harvested energy flow (power). The described very simple control algorithm can be improved so to take into consideration the possible variations of the vibration magnitude, for an appropriate adaptation of the voltages $V_1$ and $V_2$.

The proposed mixed model of the harvester is very useful for the study and the design of the system, mainly, because it relies on the physical relations between the components, in difference with SDF (signal data flow) Simulink-liked models. The observed electromechanical coupling has yet to be studied and understood in a more fundamental way.

ACKNOWLEDGMENT

This work is partially funded by the French Research Agency (ANR, Agence Nationale pour la Recherche), and is supported by the French competitiveness cluster "Ville et Mobilité Durable"